\documentclass{caosp306}

\usepackage{graphicx}
\usepackage{natbib}
\usepackage{longtable}
\articleNo{273}
\pubyear{2018}
\volume{48}
\volnumber{2}
\firstpage{319}
\received{January 16, 2018}
\accepted{February 12, 2018}

\def\BibTeX{{\rm B\kern-.05em{\sc i\kern-.025em b}\kern-.08em
             T\kern-.1667em\lower.7ex\hbox{E}\kern-.125emX}}

\begin{document}

\hauthor{M.\,Hus\'{a}rik}
\title{Shape model of the asteroid (2501) Lohja from long-term photometric observations}
\author{
        M.\,Hus\'{a}rik
       }
\institute{
           \lomnica, \email{mhusarik@ta3.sk}
          }

\date{March 8, 2003}
\maketitle

\begin{abstract}
We present lightcurves, shapes, and a 3D convex spin-axis model for the main-belt asteroid (2501) Lohja. The models were obtained with the lightcurve inversion process using combined dense photometric data from the apparitions between years 2006 and 2017. The analysis found a sidereal period of 3.80835\,hours, and a possible ecliptic pole solution (J2000.0), the prograde sense of rotation and main ratios of the ellipsoidal model.
\end{abstract}

\section{Introduction}
\label{intro}

The shape and spin axis orientation of an asteroid can be determined if a sufficient number of lightcurve data are available. For main-belt objects, data from many years (even decades), may be required to get an accurate picture, assuming that supplementary data from radar, adaptive optics, and/or occultations are not available. Distribution of pole coordinates can provide information about history in the asteroid belt and allows us to hypothesize about the changes of the rotation rates of minor planets due to thermal YORP effect.

The lightcurve database consists of the lightcurve parameters (i.e., periods and amplitudes) of more than 13\,300 minor planets (see \cite{2009Icar..202..134W} for more information about the asteroid lightcurve database), but spin vectors and shape models have only been determined for 940 of them\footnote{DAMIT database available at {\tt http://astro.troja.mff.cuni.cz/projects/damit}, for more information see the paper by \cite{Durechetal2010}}. \cite{Magnusson1989} provided a review of all techniques used in order to obtain poles and shapes of asteroids. One approach to obtain this information is to study the brightness variation of asteroids as they spin about their axes. We have used the robust method, the so-called lightcurve inversion described by \cite{KaasalainenTorppa2001} and \cite{KaasalainenTorppaMuinonen2001}. Asteroid (2501) Lohja was selected for long-term observations to obtain its precise sidereal rotational periods, the ecliptic coordinates of the rotational axis, the sense of rotation, and the convex 3D shape model. For this asteroid the estimation of the sidereal period, pole coordinates, and the 3D shape model was made for the first time.

\section{Photometry}
Asteroid (2501) Lohja was discovered on April 14, 1942, by the Finnish astronomer Liisi Oterma and it is named after the town in southern Finland. The asteroid is located in the main belt ($a=2.42$\,au, $e=0.19$, $i=3.31^\circ$), classified as an A spectral type with the diameter of about 11\,km, absolute magnitude 12.2\,mag, and albedo 0.17 (values taken from the ALCDEF database\footnote{Web page \tt http://alcdef.org}).

The first photometric observations of (2501) Lohja were obtained by \cite{Higgins2006}. Asteroid Lohja was considered to be a potentially binary system, or tumbler. Four years later more observations were made to test a duality of the system by \citet{HigginsOeyPravec2011}. They proved there was no satellite and also refined the synodic rotational period to 3.8086\,h.

The photometric observations of (2501) Lohja were performed at the Skalnat\'{e} Pleso Observatory (MPC code 056) using the 0.61-m $f/4.3$ reflecting telescope and a CCD SBIG ST-10XME camera. Observations were done in the standard Johnson-Cousins $R$ band on every scheduled observing night. We obtained CCD frames $3 \times 3$ binning and resolution of 1.6 arcsec/px, later $2 \times 2$ binning and resolution of 1.07 arcsec/px. We applied the calibration with dark and flatfield frames in the standard way. Next, we used a differential aperture photometry technique to obtain lightcurves. They were used as an input to inversion routines. The source code for these is available on the DAMIT (Database of Asteroid Models from Inversion Techniques) web page.

\section{Results}
\begin{figure}[h]
	\centerline{
		\includegraphics[width=0.94\textwidth]{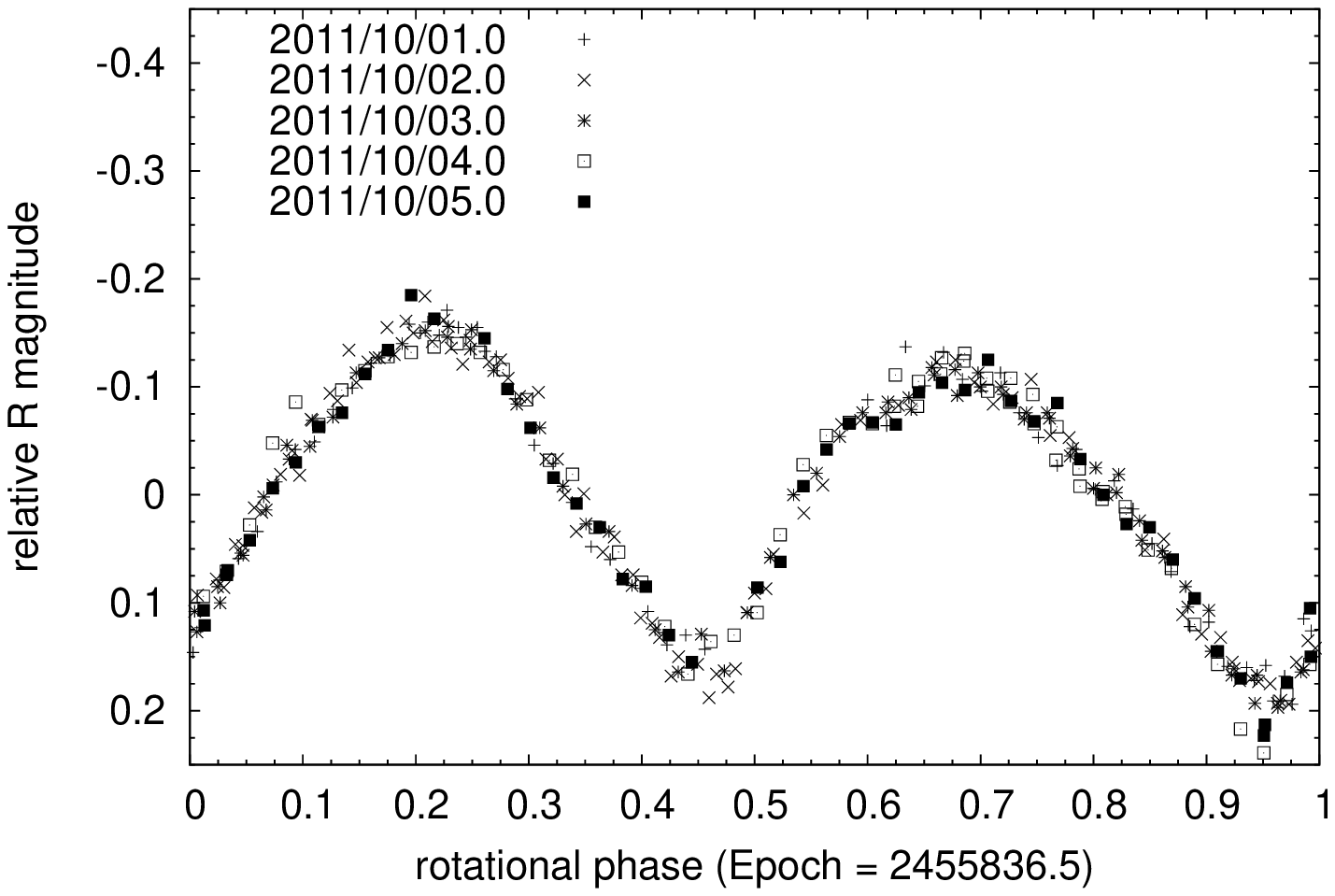}}
	\caption{Composite lightcurve of Lohja in 2011.}
	\label{Lohja_complc_2011}
\end{figure}

\begin{figure}[htb]
	\centerline{
		\includegraphics[width=0.94\textwidth]{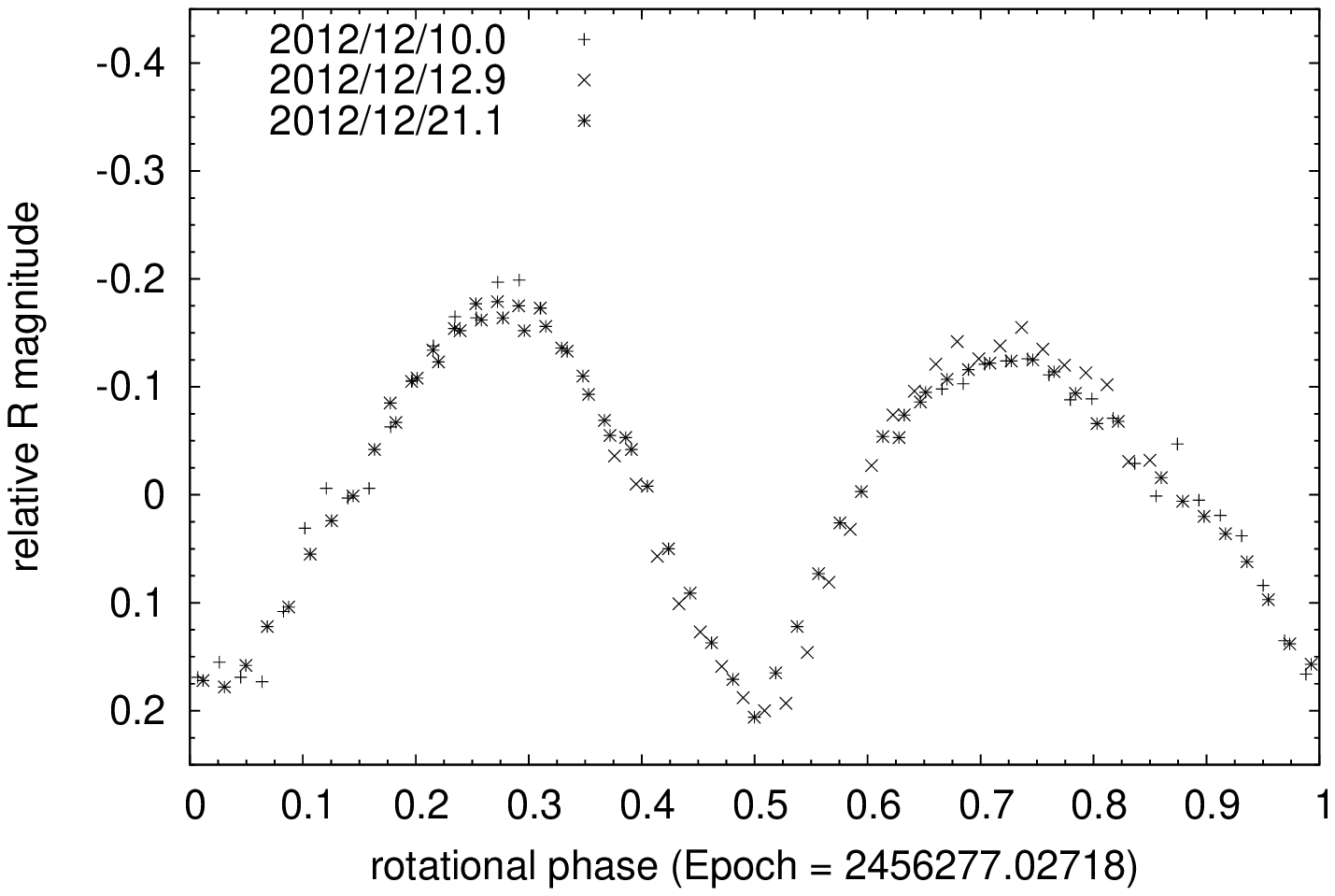}}
	\caption{Composite lightcurve of Lohja in 2012.}
	\label{Lohja_complc_2012}
\end{figure}

\begin{figure}[!htb]
	\centerline{
		\includegraphics[width=0.94\textwidth]{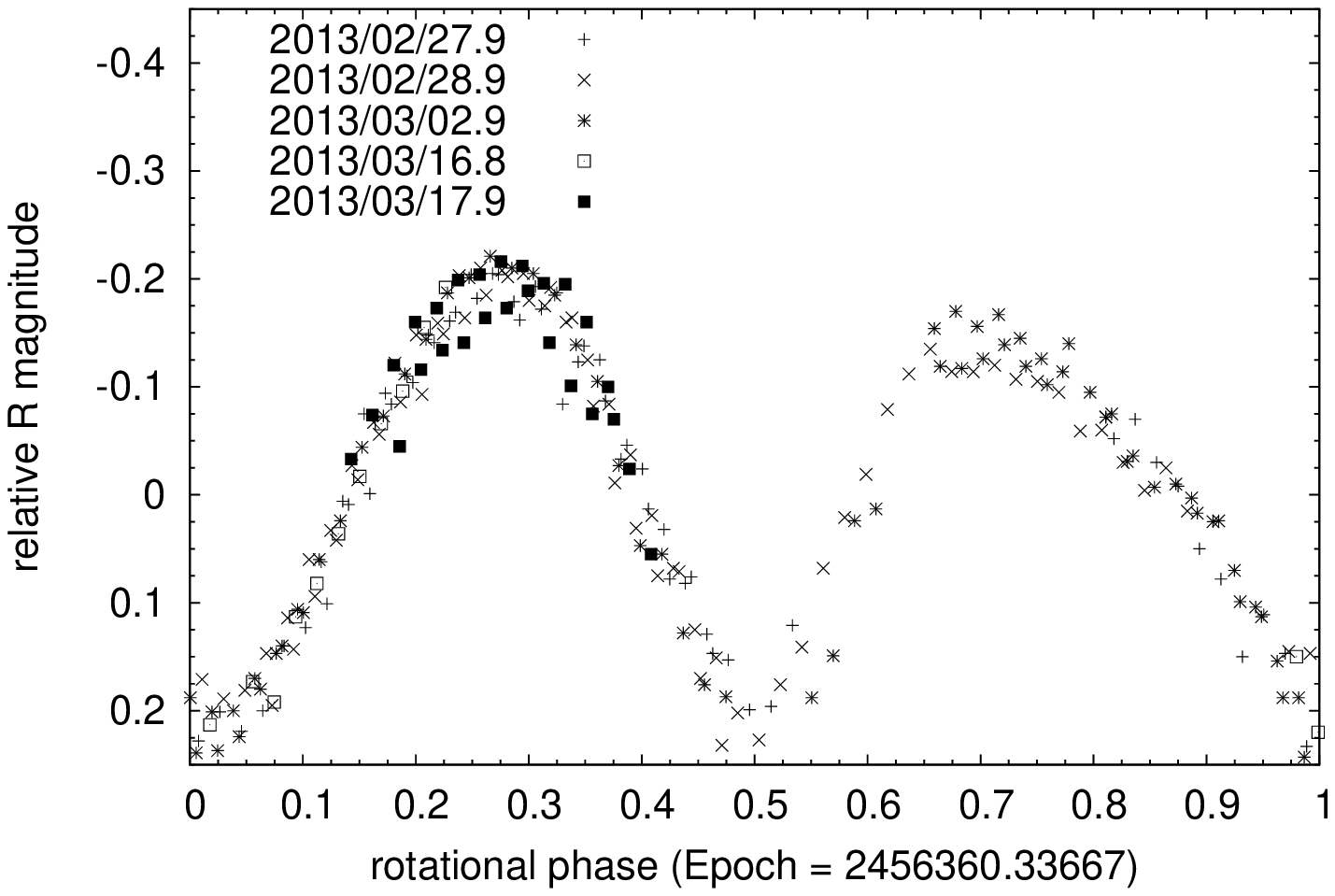}}
	\caption{Composite lightcurve of Lohja in 2013.}
	\label{Lohja_complc_2013}
\end{figure}

\begin{figure}[htb]
	\centerline{
		\includegraphics[width=0.94\textwidth]{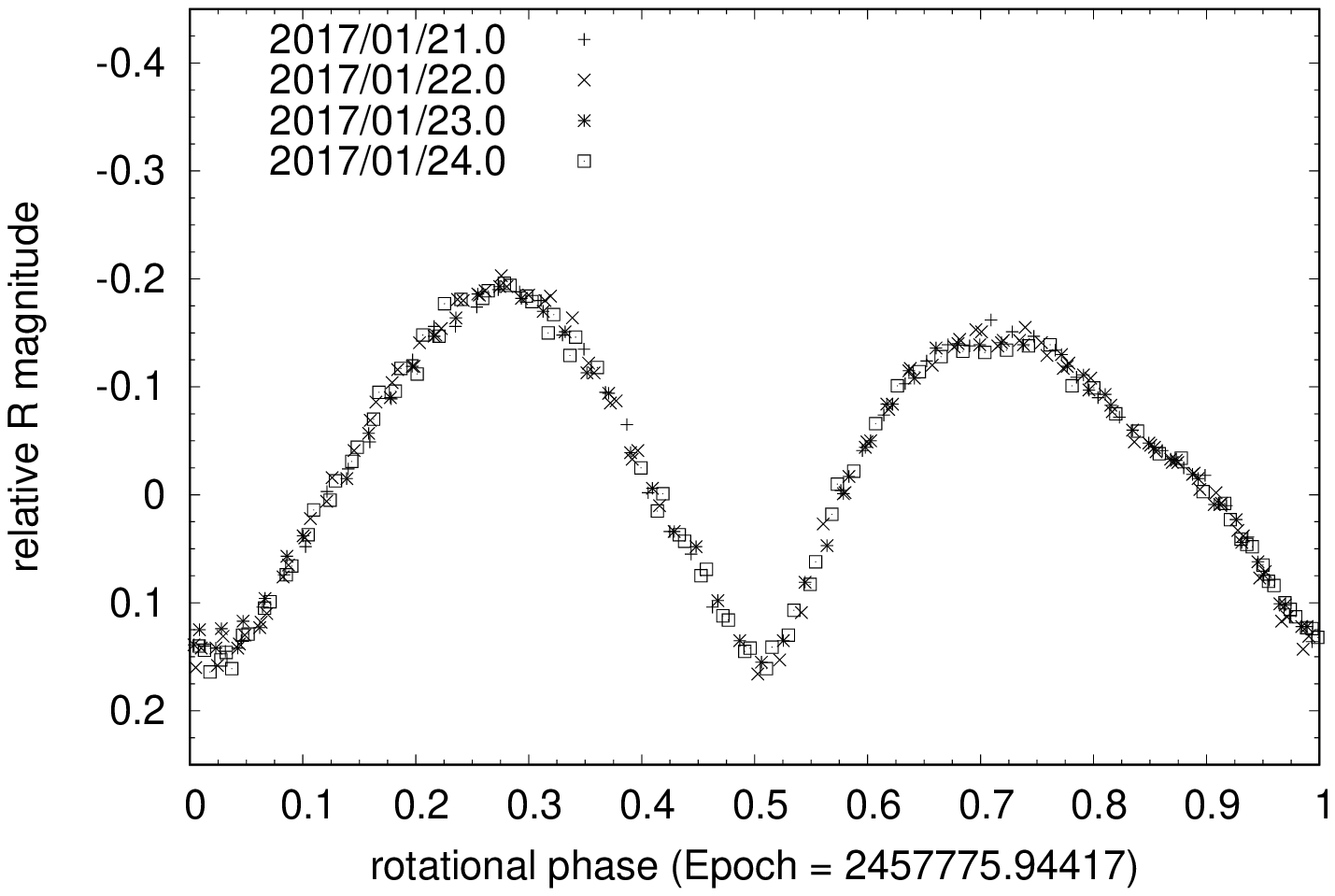}}
	\caption{Composite lightcurve of Lohja in 2017.}
	\label{Lohja_complc_2017}
\end{figure}

Our first observation of (2501) Lohja was achieved on the last day of October 2007. Due to bad atmospheric conditions the photometric run was short. In October 2011 suitable observational conditions for the Skalnat\'{e} Pleso Observatory occurred. The lightcurves from 5 nights covered the period published by \cite{HigginsOeyPravec2011}, obtaining a $3.8082 \pm 0.0003$\,h and an amplitude of 0.35\,mag (Fig.\,\ref{Lohja_complc_2011}). In the next year we observed (2501) Lohja in 3 nights and its rotational period was established at $3.8089 \pm 0.0001$\,h and an amplitude of 0.37\,mag (Fig.\,\ref{Lohja_complc_2012}). And at the beginning of the year 2013 another observations were made in 5 nights and the rotational period of $3.8084 \pm 0.0001$\,h and an amplitude of 0.41\,mag were found (Fig.\,\ref{Lohja_complc_2013}). With this package of data we performed a preliminary first shape model of Lohja presented by \cite{2014acm..conf..225H}. In 2017 there was another apparition of Lohja, and we got 7 high-quality dense photometric lightcurves (see Fig.\,\ref{Lohja_complc_2017} and corresponding data in Table\,\ref{aspect_data}). For a more precise and unambiguous model we used in our analysis another photometric data from years 2006 and 2010 available in the ALCDEF database (for more information about the photometric database see \cite{2016MPBu...43...26W}) to get a wider interval of $L_{PAB}$ and a better viewing geometry.

The first six columns of Table\,\ref{aspect_data} give the aspect data for each observing run: the mid-time of observation, heliocentric and geocentric distances, corresponding Sun-object-Earth phase angles, ecliptic coordinates of the bisector of the phase angle, data on the number of points in individual lightcurves, the total time span of the observing run, and the name of the observer or observatory.

\begin{center}
	\small
	\begin{longtable}{cccrrrcrr}
		\caption{Aspect data of (2501) Lohja. In the last column there are listed names of the observers (see Acknowledgments) and the acronym SPO for the Skalnat\'{e} Pleso Observatory.}
		\label{aspect_data}\\
		\hline \hline\noalign{\smallskip}
		Date & $r$ & $\Delta$ & $\alpha$~ & $L_{PAB}$ & $B_{PAB}$ & $N_p$ & Cov. & Obs.\\
		(UT) & (AU) & (AU) & ($^\circ$) & ($^\circ$)~ & ($^\circ$)~~ & & (h)~ &\\
		\hline
		\endfirsthead
		\caption{Continued.}\\
		\hline\noalign{\smallskip}
		Date & $r$ & $\Delta$ & $\alpha$~ & $L_{PAB}$ & $B_{PAB}$ & $N_p$ & Cov. & Obs.\\
		(UT) & (AU) & (AU) & ($^\circ$) & ($^\circ$)~ & ($^\circ$)~ & & (h) &\\
		\noalign{\smallskip}\hline\noalign{\smallskip}
		\endhead
		\hline
		\endfoot
		\endlastfoot \\
		2006 04 01.3 &  2.059 & 1.059 & 0.8 & 192.5 & $-0.1$ & 67 & 6.1&Koff\\
		2007 11 01.0 &  2.710 & 1.831 & 11.9 &   9.8 &  0.3 & 11 & 0.8&SPO\\
		2010 05 10.7 &  1.951 & 0.970 & 10.1 & 243.9 & $-3.7$ & 78 & 8.0& Higgins\\
		2010 05 15.6 &  1.951 & 0.956 & 7.4 & 244.5 & $-3.9$ & 101 & 10.1& Higgins\\
		2010 05 18.6 &  1.951 & 0.949 & 5.8 & 244.8 & $-3.9$ & 53 & 8.8& Higgins\\
		2010 06 08.6 &  1.957 & 0.959 & 8.0 & 246.8 & $-4.4$ & 56 & 7.4& Oey\\
		2010 06 09.6 &  1.957 & 0.963 & 8.6 & 246.9 & $-4.4$ & 50 & 5.5& Oey\\
		2010 06 10.6 &  1.957 & 0.966 & 9.1 & 247.0 & $-4.4$ & 76 & 7.6& Oey\\
		2010 06 10.6 &  1.957 & 0.966 & 9.1 & 247.0 & $-4.4$ & 51 & 6.7& Higgins\\
		2010 06 11.6 &  1.958 & 0.969 & 9.6 & 247.1 & $-4.4$ & 37 & 4.4& Oey\\
		2010 06 12.5 &  1.958 & 0.972 & 10.1 & 247.2 & $-4.4$ & 91 & 9.9& Oey\\
		2010 06 13.6 &  1.959 & 0.976 & 10.7 & 247.4 & $-4.4$ & 52 & 6.0& Oey\\
		2010 06 14.5 &  1.959 & 0.980 & 11.2 & 247.5 & $-4.4$ & 93 & 9.7& Oey\\
		2010 06 19.4 &  1.962 & 1.002 & 13.6 & 248.1 & $-4.4$ & 30 & 2.6& Oey\\
		2010 06 20.5 &  1.963 & 1.007 & 14.2 & 248.3 & $-4.4$ & 52 & 7.1& Oey\\
		2010 07 08.5 &  1.975 & 1.126 & 21.6 & 251.4 & $-4.4$ & 57 & 7.3& Higgins\\
		2010 07 09.4 &  1.976 & 1.134 & 21.9 & 251.6 & $-4.4$ & 21 & 2.2& Higgins\\
		2010 07 17.5 &  1.983 & 1.203 & 24.3 & 253.5 & $-4.3$ & 63 & 6.3& Oey\\
		2010 07 24.4 &  1.990 & 1.267 & 25.9 & 255.3 & $-4.3$ & 61 & 4.2& Higgins\\
		2010 07 26.5 &  1.993 & 1.288 & 26.4 & 255.9 & $-4.2$ & 69 & 7.0& Higgins\\
		2010 07 31.5 &  1.998 & 1.338 & 27.3 & 257.3 & $-4.2$ & 36 & 3.7& Oey\\
		2010 08 04.5 &  2.003 & 1.380 & 27.9 & 258.5 & $-4.1$ & 29 & 4.1& Oey\\
		2010 08 08.5 &  2.008 & 1.423 & 28.4 & 259.8 & $-4.1$ & 26 & 2.6& Oey\\
		2011 09 21.7 &  2.770 & 1.922 & 13.4 & 31.1 & 0.9 & 18 & 1.7& Oey\\
		2011 09 22.7 &  2.772 & 1.915 & 13.0 & 31.1 & 0.9 & 22 & 1.8& Oey\\
		2011 10 01.0 &  2.782 & 1.863 & 10.1 &  31.3 &  1.1 & 60 & 6.3&SPO\\
		2011 10 02.0 &  2.783 & 1.858 &  9.7 &  31.3 &  1.1 & 87 & 6.1&SPO\\
		2011 10 03.0 &  2.785 & 1.854 &  9.3 &  31.3 &  1.1 & 76 & 7.2&SPO\\
		2011 10 04.0 &  2.786 & 1.849 &  8.9 &  31.3 &  1.1 & 60 & 4.6&SPO\\
		2011 10 05.0 &  2.787 & 1.849 &  8.5 &  31.3 &  1.1 & 50 & 5.9&SPO\\ 
		2012 12 10.0 &  2.687 & 1.896 & 14.9 & 115.2 &  4.0 & 37 & 2.6&SPO\\ 
		2012 12 12.9 &  2.683 & 1.865 & 14.1 & 115.4 &  4.0 & 26 & 1.8&SPO\\ 
		2012 12 21.1 &  2.669 & 1.785 & 11.3 & 115.7 &  4.1 & 64 & 4.6&SPO\\ 
		2013 02 27.9 &  2.542 & 1.809 & 17.9 & 116.4 &  3.6 & 58 & 5.1&SPO\\ 
		2013 02 28.9 &  2.541 & 1.817 & 18.2 & 116.5 &  3.6 & 71 & 5.7&SPO\\ 
		2013 03 02.9 &  2.536 & 1.835 & 18.7 & 116.7 &  3.6 & 70 & 5.5&SPO\\ 
		2013 03 16.8 &  2.508 & 1.968 & 21.7 & 118.5 &  3.4 & 13 & 0.9&SPO\\
		2013 03 17.9 &  2.506 & 1.979 & 21.9 & 118.6 &  3.4 & 26 & 4.7&SPO\\
		2017 01 21.0 &  2.445 & 1.522 & 10.2 & 141.6 &  3.5 & 47 & 3.8&SPO\\
		2017 01 22.0 &  2.443 & 1.514 &  9.8 & 141.6 &  3.5 & 80 & 7.3&SPO\\
		2017 01 23.0 &  2.440 & 1.507 &  9.3 & 141.7 &  3.5 & 69 & 5.7&SPO\\
		2017 01 24.0 &  2.438 & 1.499 &  8.9 & 141.7 &  3.5 & 88 & 8.0&SPO\\	
		2017 03 28.9 &  2.298 & 1.569 & 20.7 & 144.1 &  2.5 & 52 & 4.6&SPO\\	
		2017 03 30.9 &  2.293 & 1.585 & 21.3 & 144.4 &  2.4 & 58 & 4.5&SPO\\	
		2017 03 31.9 &  2.291 & 1.593 & 21.6 & 144.5 &  2.4 & 45 & 4.0&SPO\\	
		\noalign{\smallskip}\hline \hline
	\end{longtable}
\end{center}


\begin{figure}[htb]
	\centerline{\includegraphics[width=0.8\textwidth]{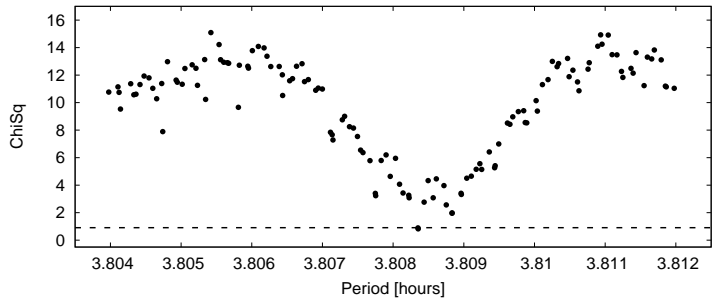}}
	\caption{A plot of $\chi^2$ vs period. the dashed line represents a value of about 10\% greater than the lowest $\chi^2$ value. One point with the lowest $\chi^2$ is the period value of 3.808\,352\,92 hours.}
	\label{PeriodSearch}
\end{figure}
\begin{figure}[ht]
	\centerline{\includegraphics[width=0.7\textwidth]{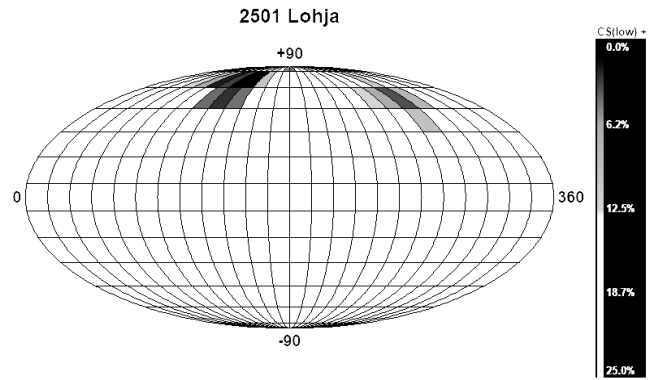}}
	\caption{$\chi^2$ residuals between the synthetic and observed lightcurves of the asteroid Lohja for spin-vector coordinates covering the entire celestial sphere. The black region represents the pole location with the lowest $\chi^2$, which increases as the color goes from dark grey to light grey and, finally, to white.}
	\label{Lohja-rms}
\end{figure} 

\begin{figure}[!h]
	\hbox{\centerline{
			\includegraphics[height=3.2cm]{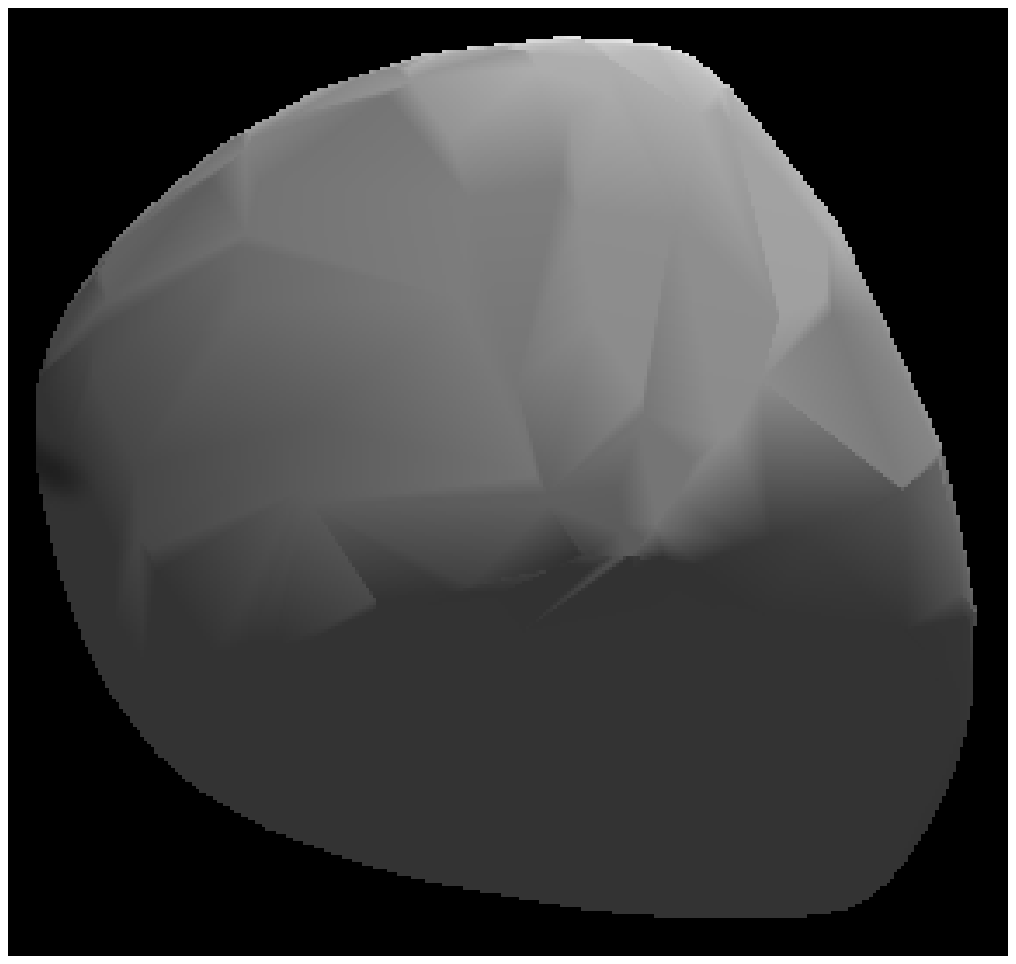}
			\includegraphics[height=3.2cm]{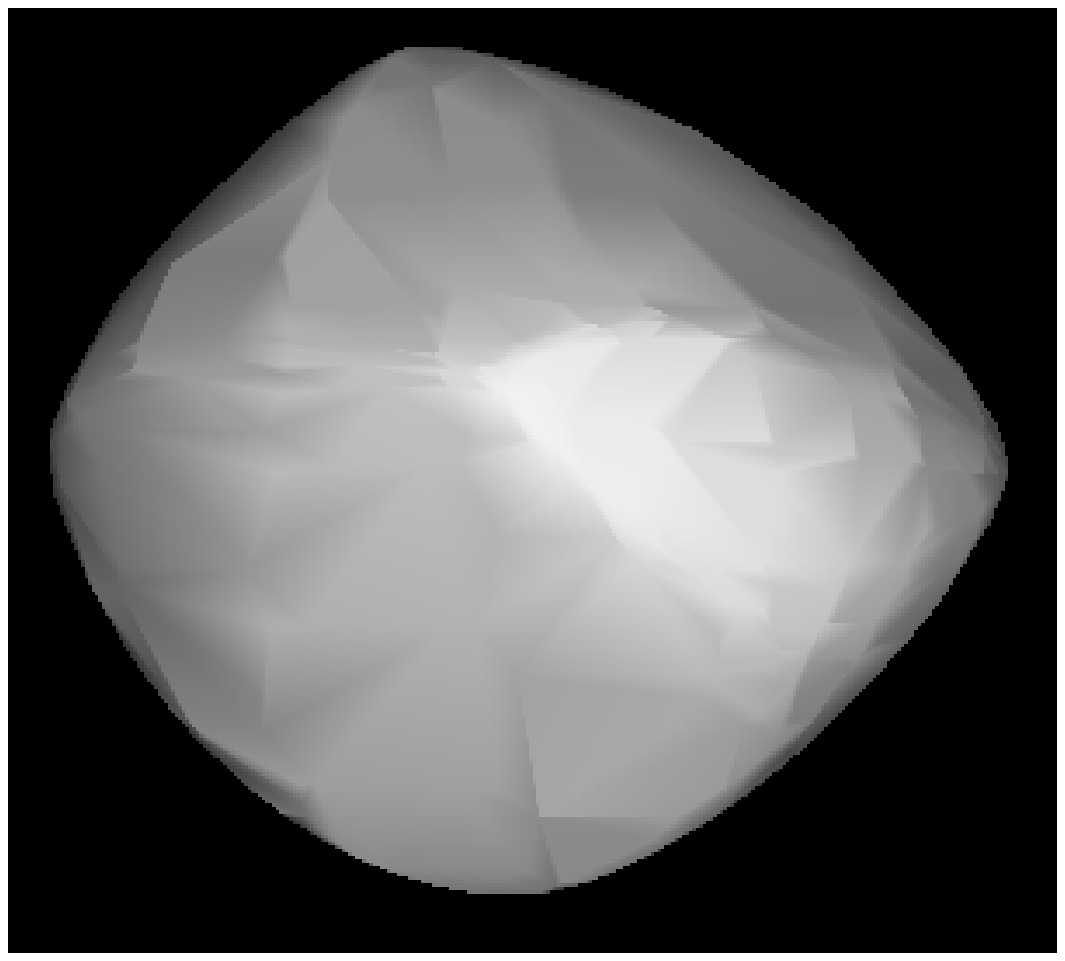}
			\includegraphics[height=3.2cm]{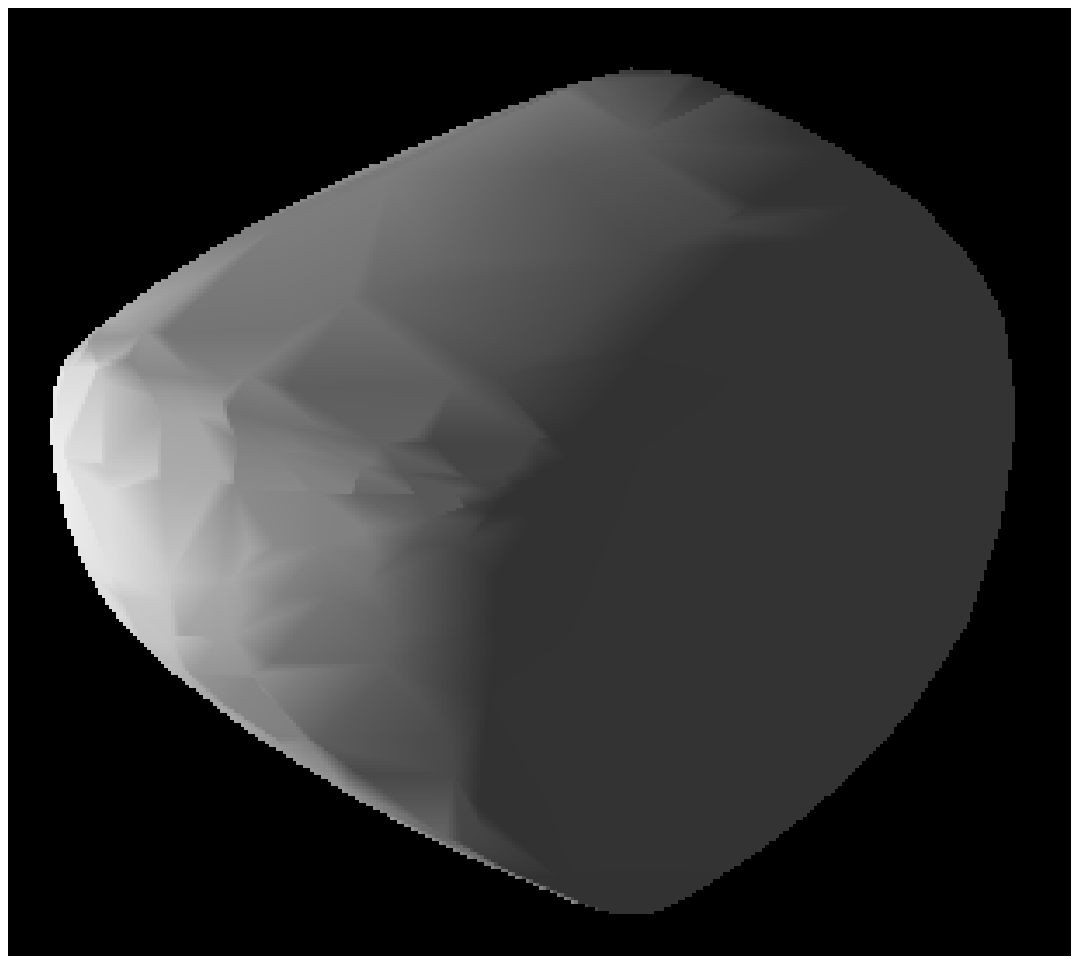}}}
	\caption{The shape model of the asteroid (2501) Lohja. On the left there is shown a north pole view, in the middle and right the equatorial viewing and illumination geometry with rotational phases $90^{\circ}$ apart. This solution is for pole determination at $(117^{\circ} \pm 10^{\circ}, +82^{\circ} \pm 10^{\circ})$.}
	\label{Lohja_3Dview}
\end{figure}  

\begin{figure}[!h]
	\hbox{\centerline{
			\includegraphics[width=0.473\textwidth]{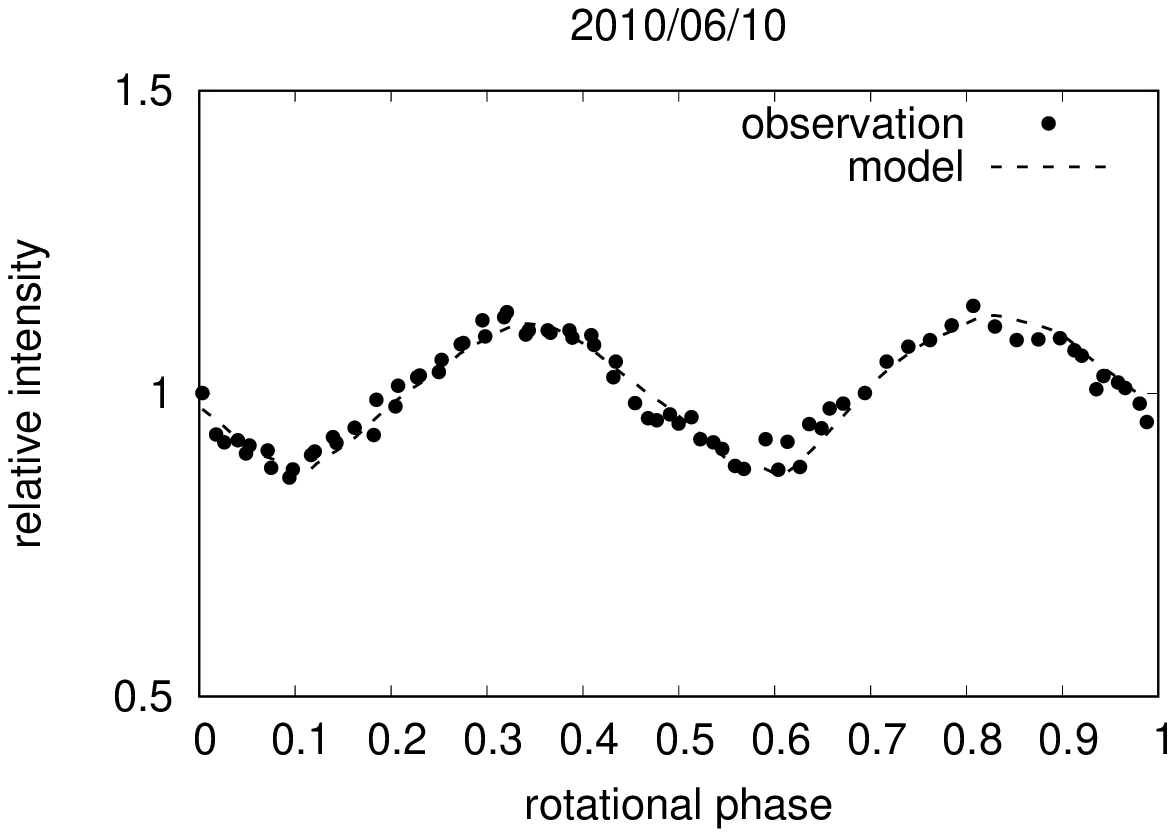}
			\includegraphics[width=0.473\textwidth]{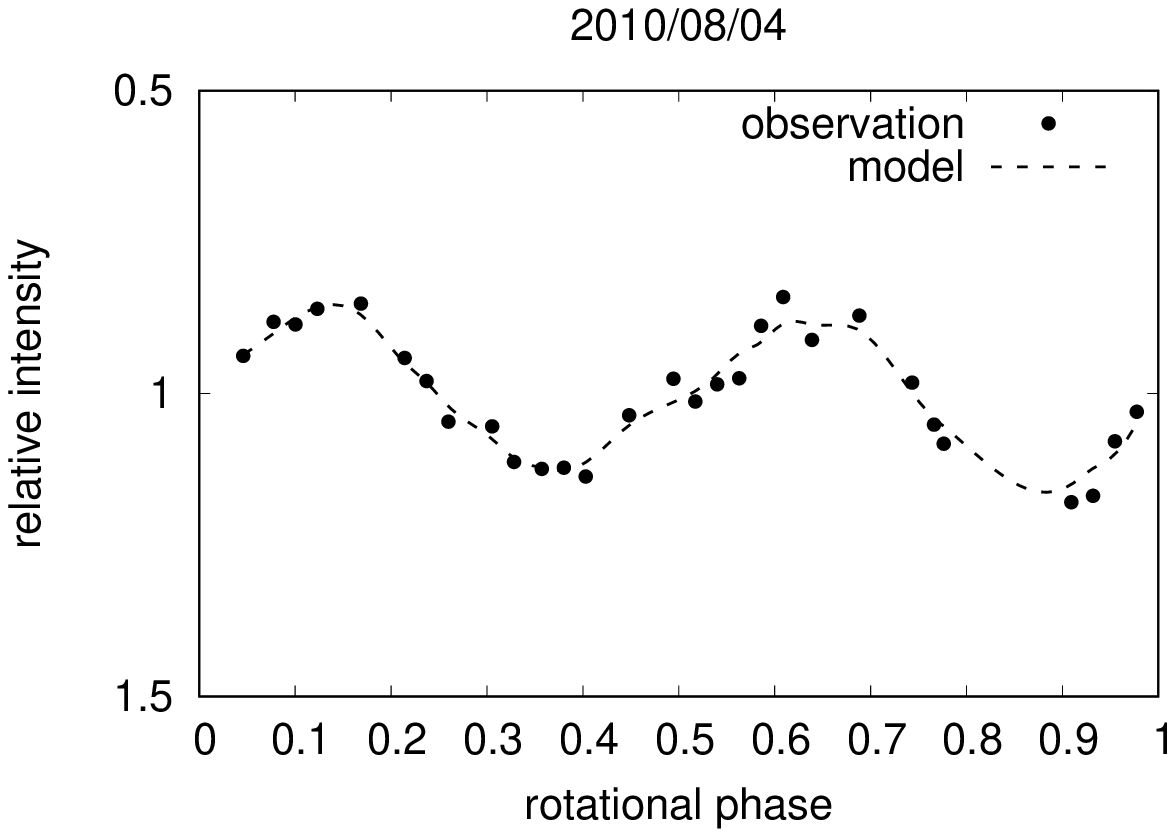}}}
	\hbox{\centerline{
			\includegraphics[width=0.473\textwidth]{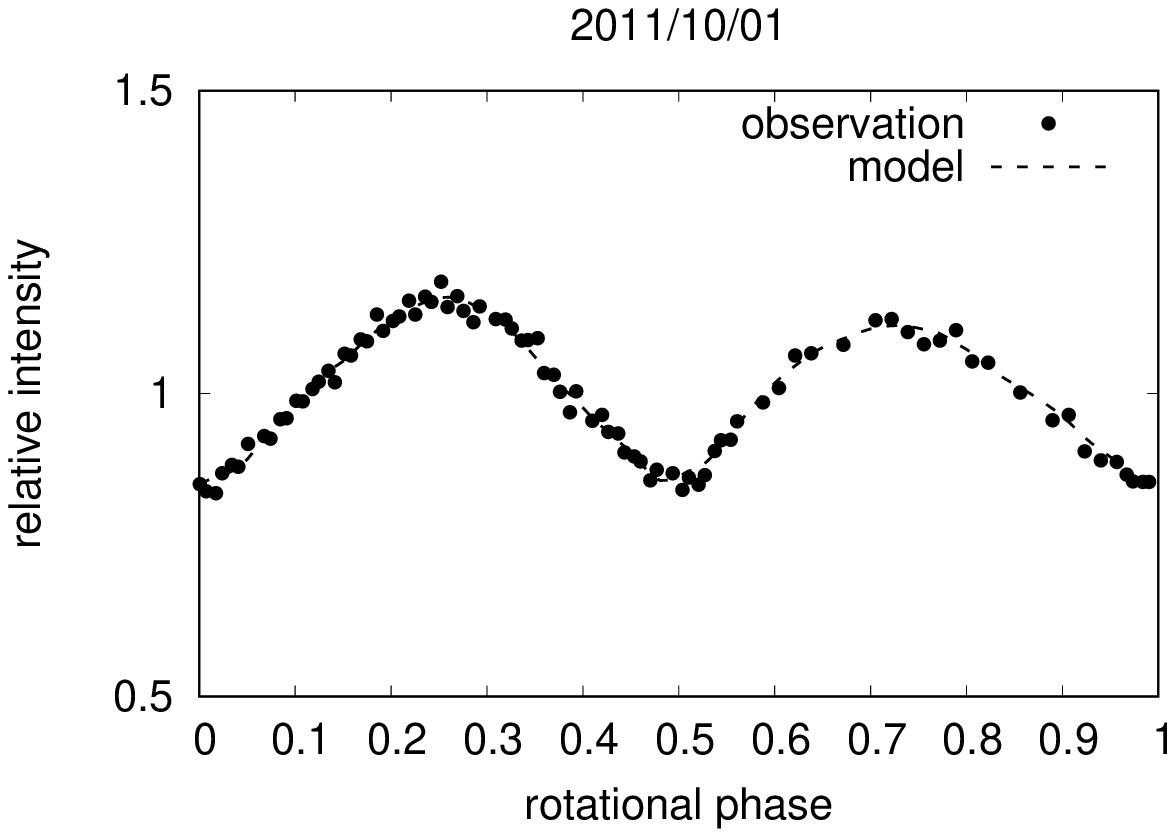}
			\includegraphics[width=0.473\textwidth]{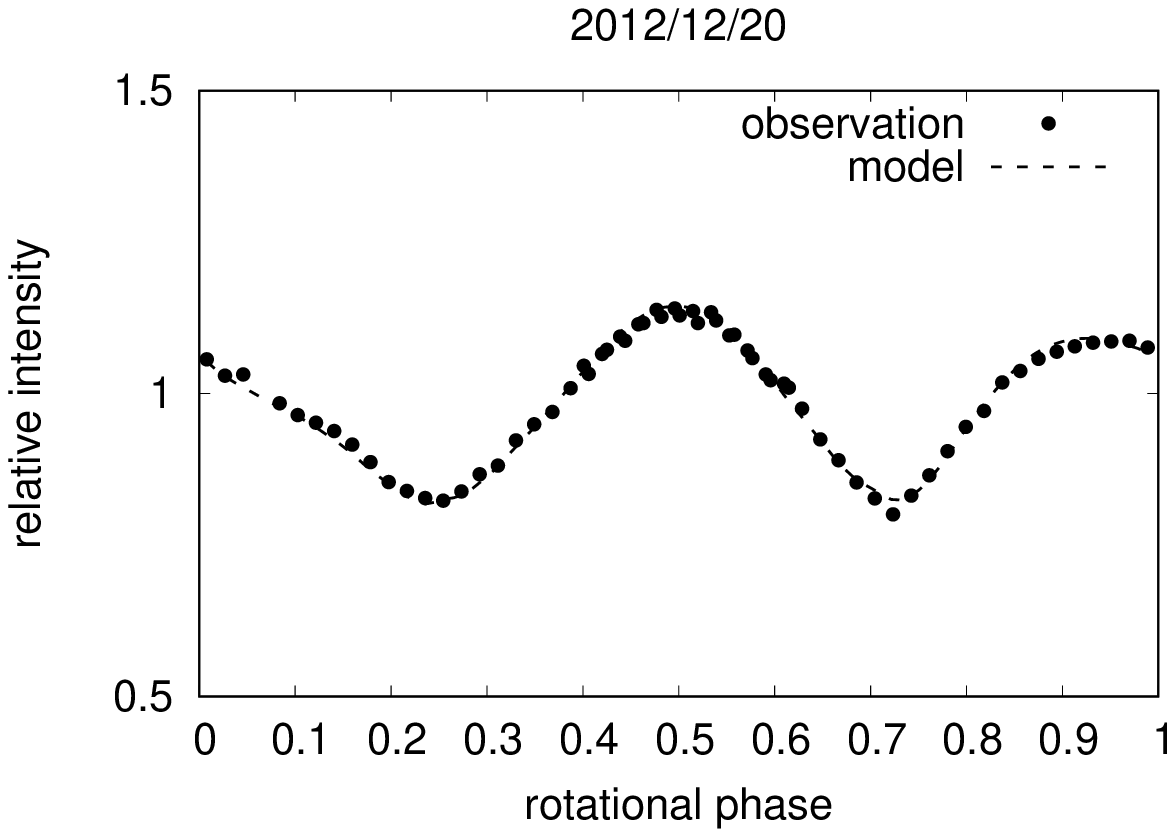}}}
	\hbox{\centerline{
			\includegraphics[width=0.473\textwidth]{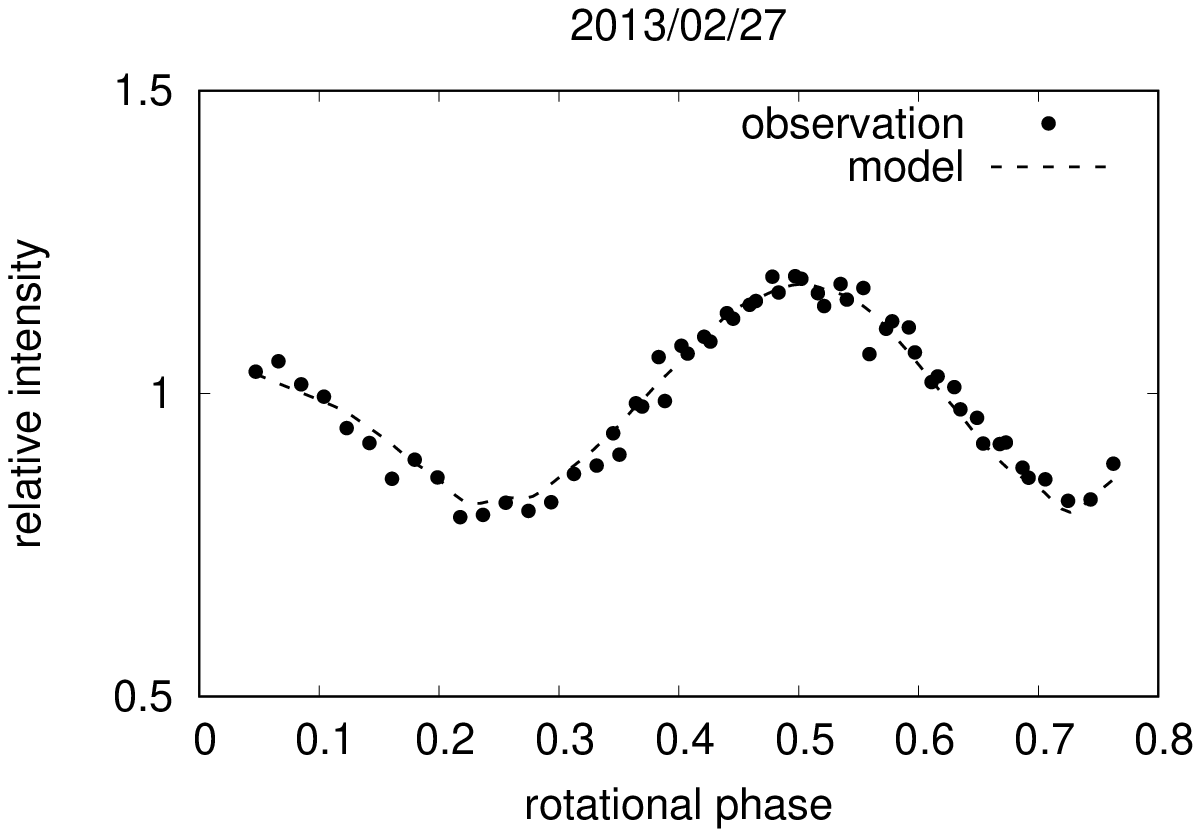}
			\includegraphics[width=0.473\textwidth]{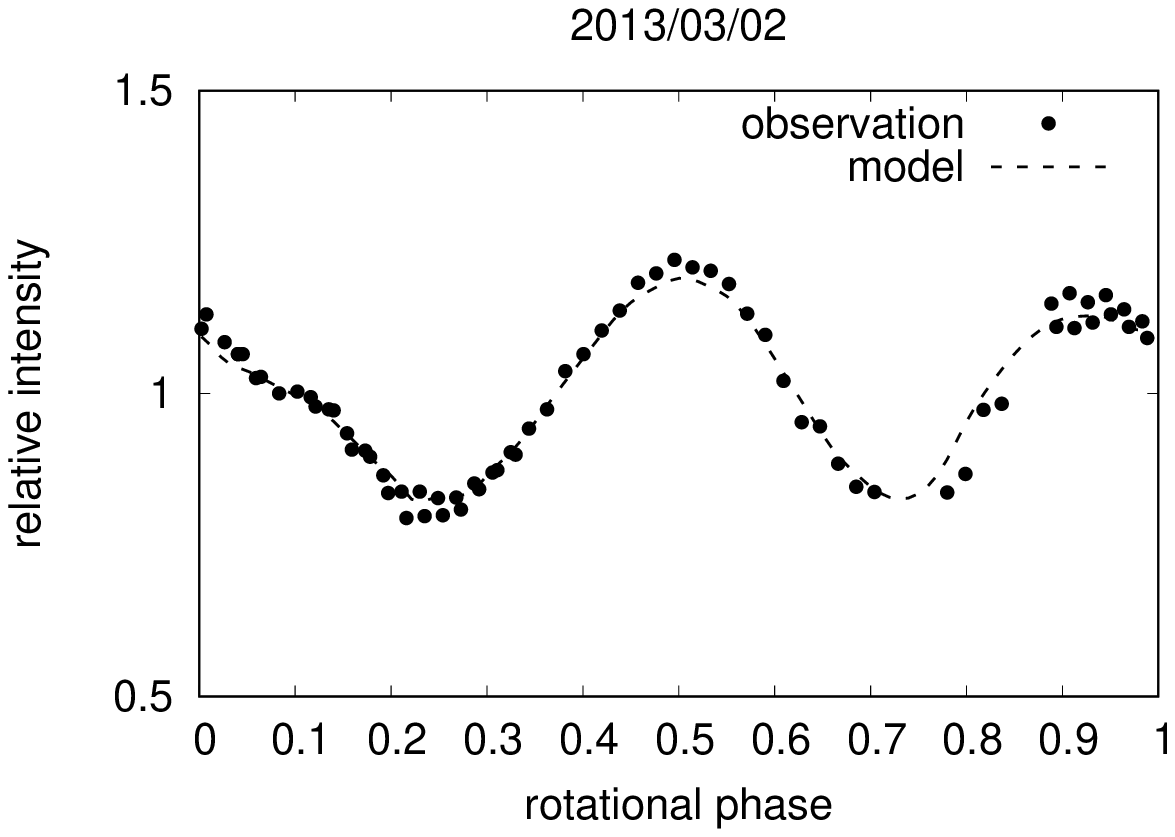}}}
	\hbox{\centerline{
			\includegraphics[width=0.45\textwidth]{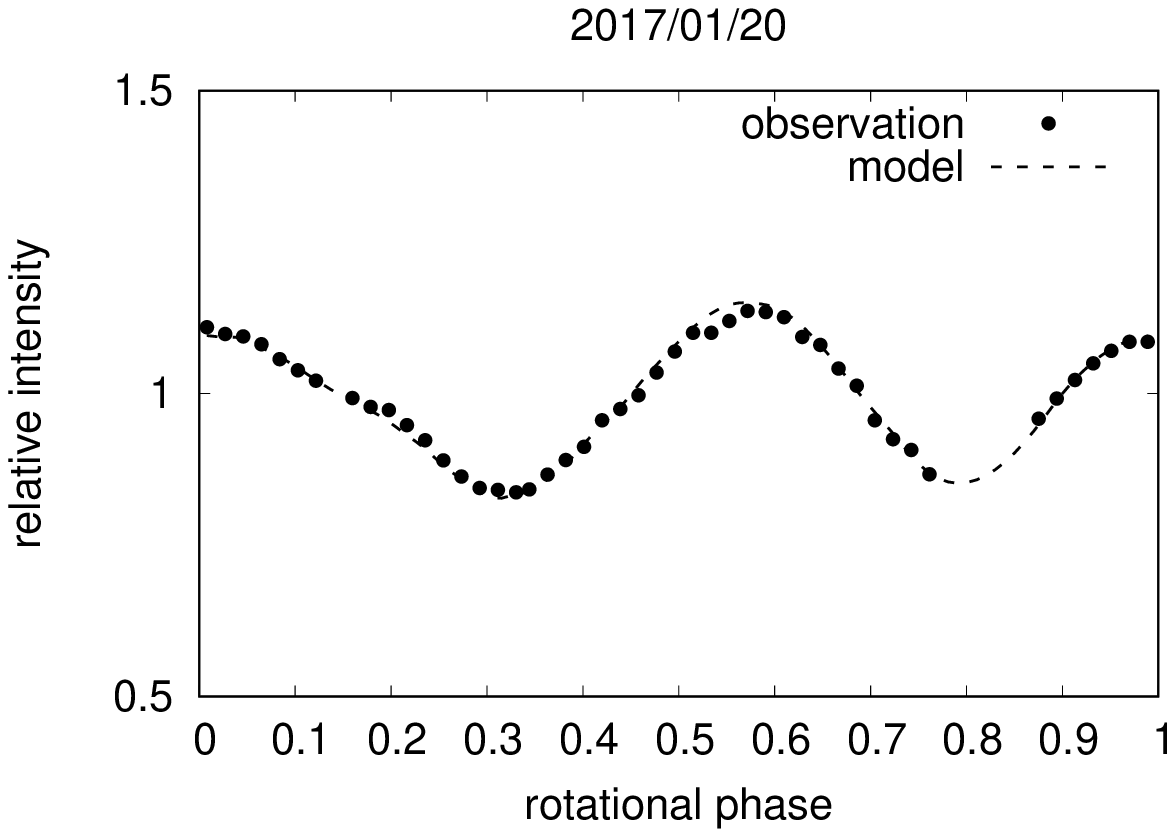}
			\includegraphics[width=0.45\textwidth]{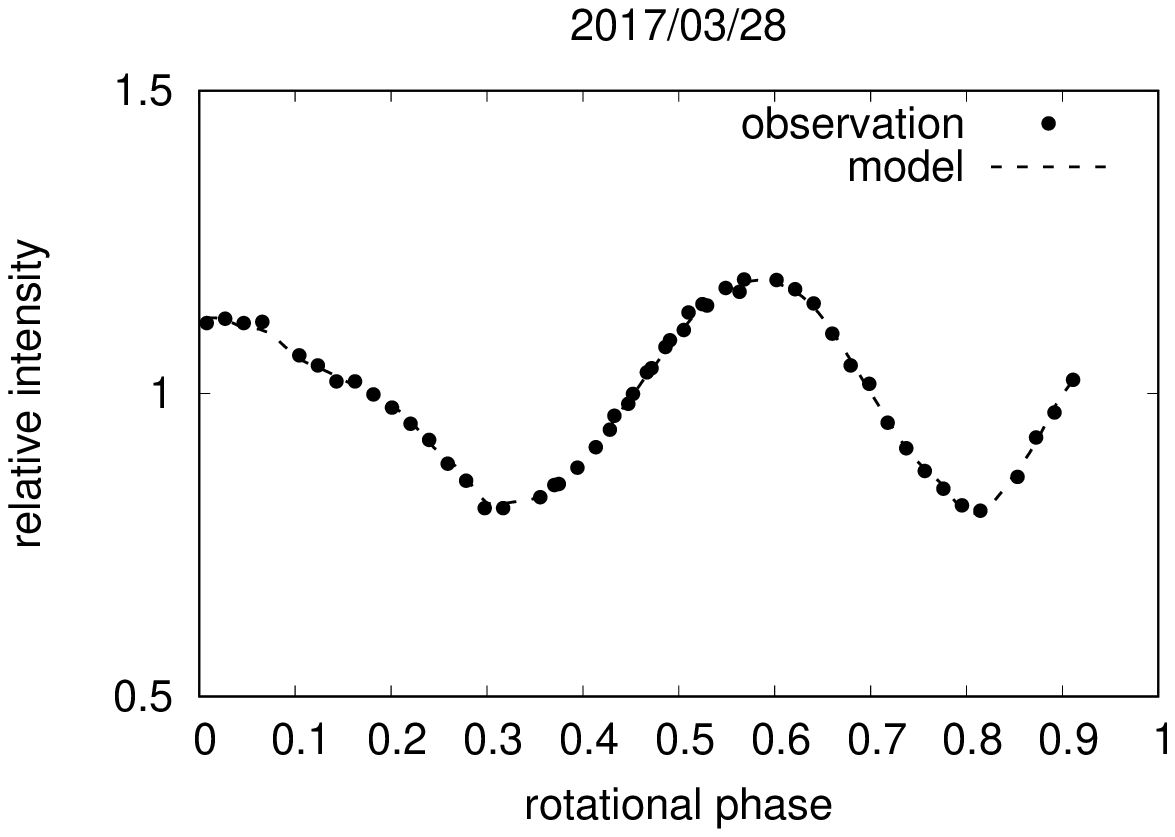}}}
	\caption{Comparison of observed data (points) and modelled brightness (the dashed curve) for eight representative lightcurves of (2501) Lohja. The plots cover one rotation cycle, the brightness is given in relative intensity units.}
	\label{Lohja_obsvsmodel}
\end{figure}

\newpage
We performed an asteroid modelling with the lightcurve inversion method described by \cite{KaasalainenTorppa2001} and \cite{KaasalainenTorppaMuinonen2001}. The method allows us to determine the sidereal period and pole solutions and recover a detailed convex 3D shape model. The model can reproduce almost all lightcurve details. The sidereal period was estimated and fixed at the value of 3.808\,352\,92 hours (Fig.\,\ref{PeriodSearch}). Next, the finding of the north pole coordinates was computed. Fig.\,\ref{Lohja-rms} shows two possible locations of the north pole, but according to an analysis of the source data the most possible position of the ecliptic longitude $\lambda_{\rm p}$ (J2000.0) and latitude $\beta_{\rm p}$ (J2000.0) is at $(117^{\circ} \pm 10^{\circ}, +82^{\circ} \pm 10^{\circ})$. The second {\em mirror} pole solution is the fact that the uncertainty cannot be removed for asteroids with orbits close to the ecliptic plane. From that it follows that the sense of rotation of asteroid Lohja is prograde. The 3D shape model has the ratios at $a/b=1.22$ and $b/c=1.10$.

The corresponding 3D shape model is presented in Fig.\,\ref{Lohja_3Dview}. To present how the model reproduces the observations, we display eight examples of observing runs in Fig.\,\ref{Lohja_obsvsmodel}.

\section{Conclusion}
We conducted detailed, long-term photometric observations of the asteroid Lohja and found synodic periods and lightcurve amplitudes at seven apparitions. We used these results to find an estimate for the ecliptic longitude and latitude of the spin axis (north pole) for that asteroid. 

Remote sensing techniques (ground-based observations) are the main source of information about the basic physical properties of asteroids such as the size, the spin state, or the spectral type. Time-resolved photometry is the most widely used method to provide data that are used to find spin states, asteroid satellites, and shapes. In the past sixteen years, asteroid lightcurve inversion has led to more than 1\,700 asteroid models. The number of asteroid models derived from lightcurves and other sources by inversion techniques is continuously growing, so we expect an increase of the number of models in the next few years with the dawn of all-sky surveys.

\acknowledgements
This article was created by the realisation of the project ITMS No.\,26220120029, based on the supporting operational Research and development program financed from the European Regional Development Fund. Scientific part of this article was realized by the bilateral project SAV-AV \v{C}R 15-17 and Grant VEGA No.\,2/0023/18. Also we want to thank R.\,A.\,Koff (Antelope Hills Observatory, USA), D.\,Higgins (Hunters Hill Observatory, Australia), and J.\,Oey (Leura Observatory, Australia) for sharing their data in the ALCDEF database.

\bibliography{mybibliography}
\end{document}